\documentclass[iop,apj,tighten]{emulateapj}
\usepackage{natbib}
\usepackage{amsmath,amstext}
\usepackage{graphicx} 
\usepackage[breaklinks,colorlinks,citecolor=blue,linkcolor=magenta]{hyperref} 
\usepackage{comment}
\usepackage{multirow}

\usepackage{amssymb}

\begin{document}
\title[]{Multiple populations in low mass globular clusters: Palomar 13
}

\author{Baitian Tang\altaffilmark{1}}
\author{Yue Wang\altaffilmark{2}}
\author{Ruoyun Huang\altaffilmark{1}}
\author{Chengyuan Li\altaffilmark{1}}
\author{Jincheng Yu\altaffilmark{1}}
\author{Doug Geisler\altaffilmark{3,4,5}}
\author{Bruno Dias\altaffilmark{6}}
\author{Jos\'e G. Fern\'andez-Trincado\altaffilmark{7}}
\author{Julio A. Carballo-Bello\altaffilmark{6}}
\author{Antonio Cabrera-Lavers\altaffilmark{8,9}}
\affil{$^1$School of Physics and Astronomy, Sun Yat-sen University, Zhuhai 519082, China; tangbt@mail.sysu.edu.cn}
\affil{$^2$Key Lab of Optical Astronomy, National Astronomical Observatories, Chinese Academy of Sciences, 20A Datun Road, Beijing 100101, China}
\affil{$^3$Departamento de Astronom\'{i}a, Casilla 160-C, Universidad de Concepci\'{o}n, Concepci\'{o}n, Chile}
\affil{$^4$Instituto de Investigaci\'on Multidisciplinario en Ciencia y Tecnolog\'ia, Universidad de La
Serena. Avenida Ra\'ul Bitr\'an S/N, La Serena, Chile}
\affil{$^5$Departamento de Astronom\'ia, Facultad de Ciencias, Universidad de La Serena. Av.
Juan Cisternas 1200, La Serena, Chile}
\affil{$^6$Instituto de Alta Investigaci\'on, Universidad de Tarapac\'a, Casilla 7D, Arica, Chile}
\affil{$^{7}$Instituto de Astronom\'ia y Ciencias Planetarias, Universidad de Atacama, Copayapu 485, Copiap\'o, Chile.}
\affil{$^8$GRANTECAN, Cuesta de San Jos\'e s/n, E-38712 , Bre\~na Baja, La Palma, Spain}
\affil{$^9$Instituto de Astrof\'isica de Canarias, V\'ia L\'actea s/n, E38200, La Laguna, Tenerife, Spain}

\begin{abstract}
Since the discovery of chemically peculiar stars in globular clusters in the last century, the study of multiple populations has become increasingly important, given that chemical inhomogeneity is found in almost all globular clusters. Despite various proposed theories attempting to explain this phenomenon, fitting all the observational evidence in globular clusters with one single theory remains notoriously difficult and currently unsuccessful. In order to improve existing models and motivate new ones, we are observing globular clusters at critical conditions, e.g., metal-rich end, metal-poor end, and low mass end. In this paper, we present our first attempt to investigate multiple populations in low mass globular clusters. We obtained low-resolution spectra around 4000 \AA~of 30 members of the globular cluster Palomar 13  using OSIRIS/Multi-object spectrograph mounted at the Gran Telescopio Canarias. The membership of red giant branch stars is confirmed by the latest proper motions from Gaia DR2 and literature velocities. After comparing the measured CN and CH spectral indices with those of the stellar models, we found a clear sign of nitrogen variation among the red giant branch stars. Palomar 13 may be the lowest mass globular cluster showing multiple populations. 

\end{abstract}

\maketitle

\section{Introduction}
\label{sect:intro}

Once prototypical  simple stellar populations, globular clusters (GCs) are now found to host multiple populations (MPs) using both photometry and spectroscopy \citep[e.g.,][]{Piotto2015, Milone2015, Carretta2010b, Meszaros2020}. Chemical abundances from spectroscopic data suggest that GCs have a group of enriched stars with enhanced N, Na, (sometimes He, Al and Si), but depleted C, O, (sometimes Mg). The observed (anti-)correlations between these light elements in GC stars indicate possible nuclear chains, e.g., \citet{Arnould1999} clearly illustrated the CNO, NeNa, MgAl cycles and the possible Si leakage with element abundance calculations using updated thermonuclear rates. Several possible astrophysical sites were proposed, e.g., AGB stars experiencing hot-bottom burning \citep{DErcole2008,DErcole2010}, fast rotating massive stars \citep{Decressin2007}, very massive stars \citep[$\sim10^4$ M$_{\odot}$;][]{Denissenkov2014}, and winds of massive stars and AGB \citep{Kim2018}. However, all of the proposed theories still face critical problems when matching the observation \citep{Bastian2018}. 

Observations of GCs at critical conditions are touchstones to examine existing models and motivate new ones (e.g., the metal-rich end, \citealt{Tang2017}; the metal-poor end, \citealt{Tang2018}). In this sense, the proposed critical GC mass which separates multiple from single populations is an important parameter for modelling MPs. \citet{Carretta2010b} proposed 4$\times$10$^4$ M$_{\odot}$~(M$_V\sim-5$ mag) as the mass limit that separates globular and open clusters (their Figure 3), suggesting this is the minimum mass required by a cluster to retain ejecta from an initial generation and allow the formation of a subsequent generation. This possible lower mass limit is recently investigated by several groups. For example, \citet{Bragaglia2017} suggested that NGC 6535 (M$_V\sim-4.75$ mag, mass$\sim2.21\times 10^4$ M$_{\odot}$) is currently the lowest mass GC that exhibits MPs using high-resolution spectroscopy (VLT-FLAMES).  While high-resolution spectra give more information for a variety of elements, it takes more time to reach enough signal to noise ratio (SNR) for fainter stars, thus limiting the sample size when studying low mass and remote GCs. Alternatively, low-resolution spectra around 4000 \AA~help to identify MPs with the well known CN-CH features, while maintaining reasonable SNR for faint stars.
Toward this end, the absence of MPs in E3 (M$_V\sim-4.12$ mag, mass$\sim1.4\times 10^4$ M$_{\odot}$, \citealt{Salinas2015})\footnote{The absence of MPs in E3 was later supported by high-resolution spectra  analysis of four member stars \citep{Monaco2018}} and the probable presence of MPs in ESO452-SC11 (M$_V\sim-4.02$ mag, mass$\sim (6.8\pm3.4)\times 10^3$ M$_{\odot}$, \citealt{Simpson2017}) push this possible threshold mass to a lower value. However, whether the threshold mass exists is still uncertain.


To further explore this possible threshold mass, we have obtained low-resolution spectra for the members of several remote, low mass GCs which deserve new or updated observation and investigation, especially in terms of MPs. In this first paper, we will explore the better studied GC among our sample: Palomar 13 (hereafter Pal 13). The globular cluster Pal 13 is one of the faint, low mass GCs identified by the Palomar Observatory Sky Survey \citep{Abell1955}. It is located away from the Galactic plane ($\alpha_{\rm J2000}=23^h 06^m 44^s .48$, $\delta_{\rm J2000}=+12^{\circ} 46' 19.2''$, $l=87^{\circ}.10379$, $b=-42^{\circ}.69993$)\footnote{\citet{DC2006}}, with a Galactocentric distance of 25 kpc \citep[][2010 edition]{Harris1996}, placing it well into the halo of the Milky Way (MW). Although the Harris catalog suggests M$_V\sim -3.7$ mag for Pal 13, recent deep photometric survey by \citet{Munoz2018} shows that M$_V$ should be $-2.84\pm0.55$ mag, which is the faintest GC (in terms of $V$ band absolute magnitude) among other MP-investigated GCs. Precise proper motion (PM) was first derived by \citet{Siegel2001} using moderate-scale photographic plates separated by a 40 year baseline, ($\mu_{\alpha cos \delta}, \mu_{\delta}$)=(2.30, 0.27)$\pm$(0.26,0.25) mas yr$^{-1}$. This value generally agrees with the recent Gaia DR2 results,  ($\mu_{\alpha cos\delta}$, $\mu_{\delta}$)=(1.625,0.114) $\pm$ (0.076, 0.059) mas yr$^{-1}$  \citep{Vasiliev2019}. \citet[][hereafter B11]{Bradford2011} presented a detailed photometric and spectroscopic study of Pal 13 using Canada-France-Hawaii Telescope/MegaCam and Keck/DEIMOS. They derived a GC radial velocity (RV) of 26.2$\pm$2.2 km s$^{-1}$, and a metallicity of [Fe/H$]=-1.6\pm0.1$. The metallicity of Pal 13 has been debated since its discovery, with literature vales varying between $-1.5$ and $-1.9$ (e.g., \citealt{Zinn1984}; \citealt{Friel1982}). \cite{Koch2019} determined the chemical abundances based on the coaddition of the high-resolution spectra of 18 Pal 13 members. They determined the metallicity to be $-1.91\pm0.05$ (statistical) $\pm0.22$ (systematic).
Pal 13 is also one of the lowest mass GCs. B11 found the dynamical mass inside the half-light radius to be $1.3^{+2.7}_{-1.3} \times 10^3$ M$_{\odot}$. More interestingly, this GC is possibly under dissolution \citep{Piatti2020, Shipp2020}. We carefully discuss its implications in Section \ref{sect:dis}. The low mass of Pal 13 is the reason why we selected this GC for investigating MPs in the low mass end.
  
This paper is organized as follows: We describe our observation and major procedures of data reduction in Section \ref{sect:obs}. The implications of our discovery in the context of MPs are discussed in Section \ref{sect:dis}. Finally, we present a quick summary and a brief look to the future in Section \ref{sect:con}.

\begin{figure*}
\centering
\includegraphics [width=0.7\textwidth]{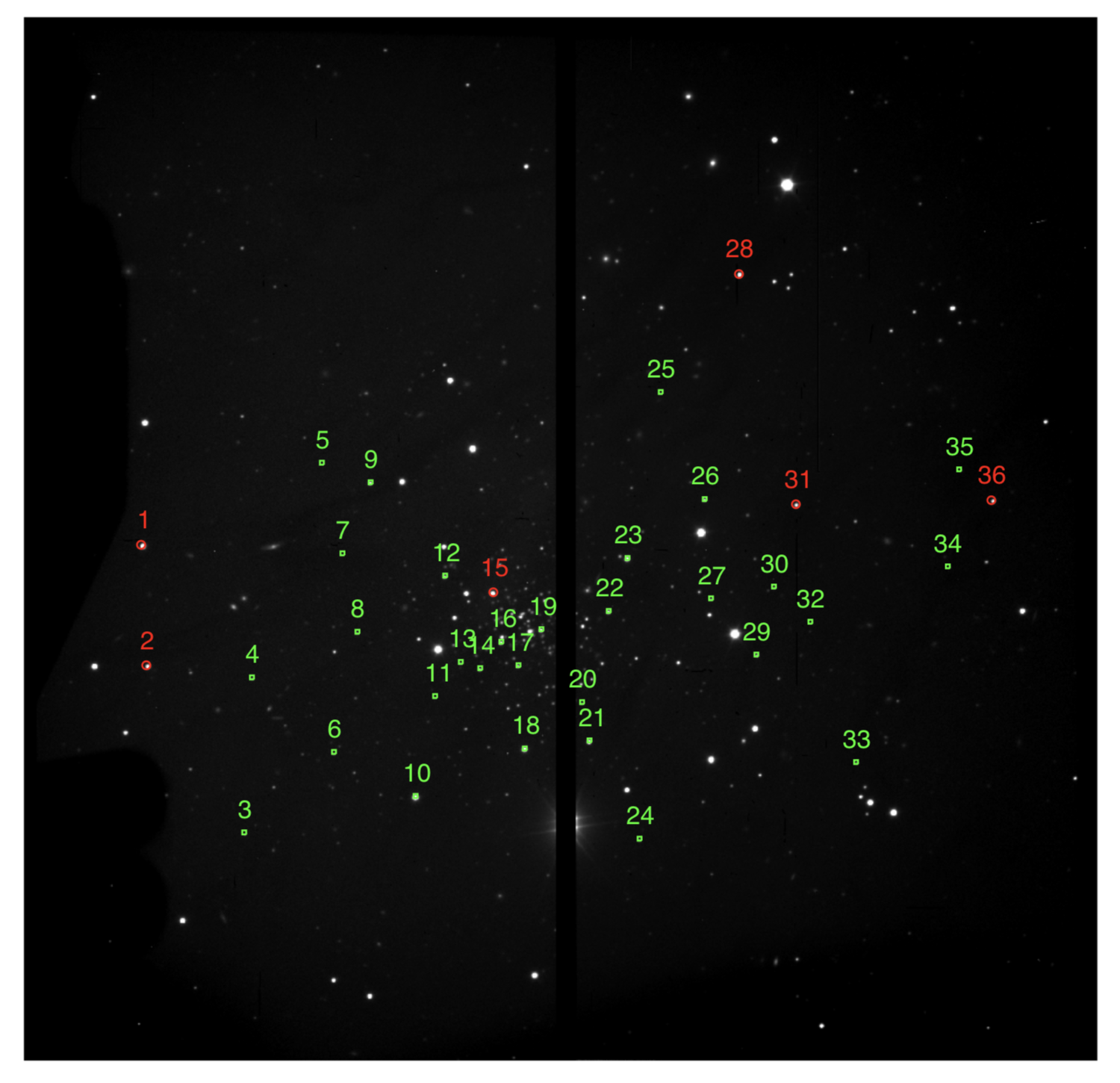} 
\caption{Locations of the observed targets and slitlets in the field of Palomar 13. Science targets are marked by green rectangular slitlets, and fiducial stars used for plate alignment are marked by red circular slitlets. The background image was taken using MOS image mode in $r$-band for 30 seconds.
}\label{fig:slit}
\end{figure*}

\section{Observation and Data Reduction}
\label{sect:obs}

\subsection{Observation}
The observations were carried out using OSIRIS/Multi-object spectrograph (MOS)\footnote{http://www.gtc.iac.es/instruments/osiris/osiris.php} mounted at the Gran Telescopio Canarias (GTC), Observatorio del Roque de los Muchachos. We first selected science targets based on RV-selected members from B11. We further considered two issues before finalizing the observed sample: 1. Slitlet observation forbids two targets from overlapping in the dispersion direction; 2. Observing brighter targets generates spectra with higher signal-to-noise ratio (SNR).  The locations of the observed targets are illustrated in Figure \ref{fig:slit}. Among them, 30 stars are science targets (green rectangular slitlets), while 6 are fiducial stars (red circular slitlets) used for plate alignment. Among the science targets, 28 were selected from the ``clean member sample'' of B11, which show consistent PMs based on the work of \citet{Siegel2001}; two targets were selected from the ``full member sample'', where their PM information is either missing or inconsistent with being cluster members. Our spectral observation roughly covers the U band (3500-4600 \AA) with nominal resolution $R\sim2500$. The observation was carried out under the program GTC2-18BCNT in December 2018, where 5 hours divided into 10 exposures of 1800s each were obtained using the OSIRIS MOS configuration. 

\subsection{Data Reduction}

\begin{figure}
\centering
\includegraphics [width=0.5\textwidth]{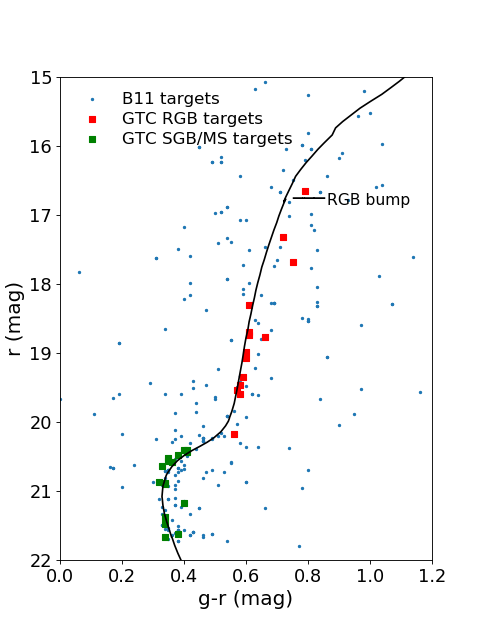} 
\caption{Color-magnitude diagram of Pal 13. The sample of \citet{Bradford2011} is shown as small dots, while our RGB (SGB/MS) science targets are labeled as red (green) squares.  PARSEC isochrone of 12 Gyr age and [Fe/H$]=-1.6$ is shown, and the location of RGB bump is labeled out.
}\label{fig:CMD}
\end{figure}

\begin{figure}
\centering
\includegraphics [width=0.5\textwidth]{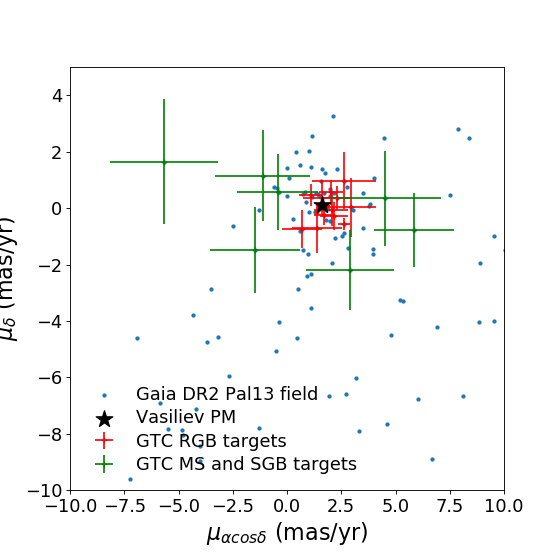} 
\caption{Proper motions (PMs) of stars near the field of Pal 13. Background stars are shown as small dots. Brighter RGB stars (Figure \ref{fig:CMD}) are marked as red error bars, while fainter SGB/MS stars are labeled as green error bars. The uncertainties of PMs are indicated by the size of the error bars. The mean PM of Pal 13 published by \citet{Vasiliev2019} is shown as black star symbol.
}\label{fig:pm}
\end{figure}

The color-magnitude diagram (CMD) of our observed science targets is shown in Figure \ref{fig:CMD}. Among them, there are 14 red giant branch stars (RGBs) and 16 sub-giant branch or main sequence stars (SGBs/MSs). The PMs of our sample stars were partially checked by B11. Here we re-checked their proper motions using the latest Gaia DR2 (Figure \ref{fig:pm}). We present the RGB subsample with red symbols, and the SGB/MS subsample with green symbols. Clearly, stars in the RGB subsample are brighter, and thus have smaller PM errors. On the other hand, stars in the SGB/MS subsample are fainter --- most of them have Gaia $G$ band magnitude greater than 20 mag, where PM becomes less trust-worthy\footnote{The typical uncertainty of stars with $20<G<21$ mag is $1.2-3$ mas/year \citep{Brown2018}.}. Therefore, the SGB/MS subsample stars either have no PM measurements in Gaia DR2 or have very large PM errors. Figure \ref{fig:pm} indicates that all of our 14 RGB stars are PM members, in agreement with B11, and the PM measurements of individual stars are consistent with that of the GC mean by \citet{Vasiliev2019}, ($\mu_{\alpha cos\delta}$, $\mu_{\delta}$)=(1.625,0.114) (mas/year). 

We used the GTCMOS pipeline developed by Dr. Divakara Mayya\footnote{https://www.inaoep.mx/$\sim$ydm/gtcmos/gtcmos.html}. We refer the readers to \cite{GG2016} for detailed description of the pipeline procedure. Given that the observations were taken with five discrete times/blocks, with each block consisting of two exposures of 1800s each, we reduce the spectra as follows: For each observation block, the observed spectral images were bias-subtracted, combined, wavelength calibrated, flat-field corrected, and extracted to multiple one-dimension (1d) spectra using the GTCMOS pipeline.  After that, we used the Ca HK feature to de-redshift the spectra to the rest frame. Since our estimation of CN and CH spectral indices partially depends on the spectral shape, we carefully calibrated the observed 1d spectra based on their stellar model spectra in the following steps. This calibration procedure was partially inspired by the work of \citet{Gerber2020}. In order to generate model spectra, stellar parameters are required. We used de-reddened $(B-V)$ colors to calculate effective temperatures of our science targets, adopting E$(B-V) =$ 0.11 \citep{Schlegel1998}. Stetson photometry provides $(B-V)$ colors for 13 brighter stars. For fainter ones, the de-reddened $(B-V)$ colors were converted from de-reddened $(g-r)$ colors provided by B11. The color-color conversion is based on the work of \citet{Jester2005}. 
The relation between $(B-V)$ colors and T$_{\rm eff}$ was inferred from the PARSEC isochrone of 12 Gyr and [Fe/H$]=-1.6$ (B11 suggested values). We the estimated $ \log(g)$ from the same isochrone using the calculated T$_{\rm eff}$. 

\begin{figure*}
\centering
\includegraphics [width=\textwidth]{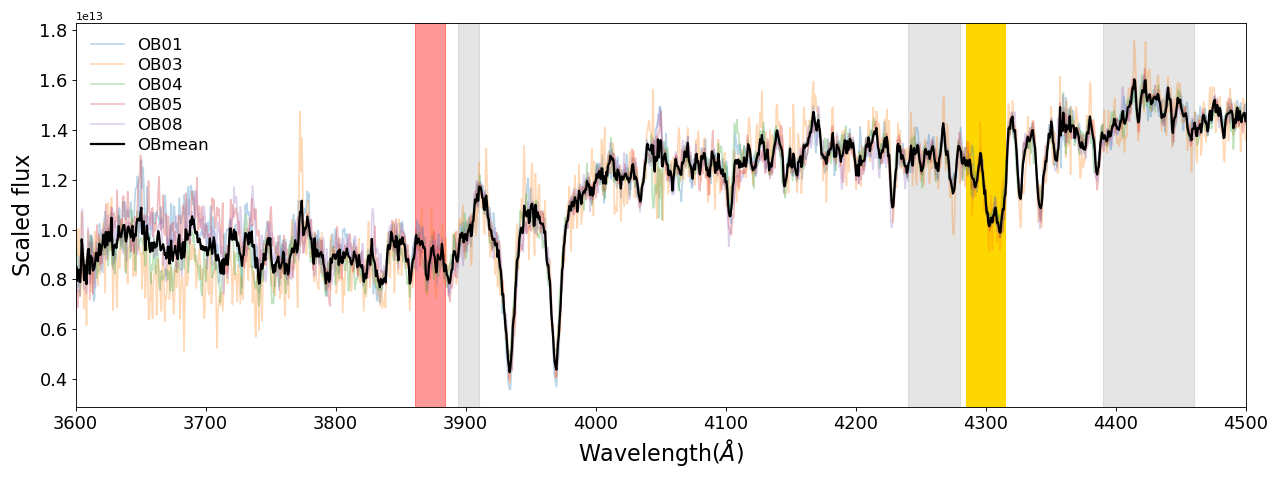} 
\caption{An example of the calibrated spectra. OB01, OB03, OB04, OB05, OB08 represent the calibrated spectra of different observation blocks, and OBmean (black line) is the mean-combined spectrum of the five observation blocks.  The spectral index features of CN3839 and CH4300 are shown in red and yellow, respectively. The spectral index pseudo-continuum are shown in gray.
}\label{fig:10}
\end{figure*}

Next, We used the estimated stellar parameters with solar-scaled abundances to generate the synthetic spectrum for each star using iSpec \citep{BC2014}\footnote{https://www.blancocuaresma.com/s/iSpec}. 
The radiative transfer code and the line lists come from SPECTRUM\footnote{http://www.appstate.edu/$\sim$grayro/spectrum/spectrum.html}. The MARCS model atmospheres \citep{Gustafsson2008} and \citet{Asplund2005} solar abundances are adopted. 
The continuum regions of spectrum of each observation block was first fitted with 3rd or 4th order polynomials, and then calibrated to match the shape of the synthetic spectra. Our definition of the continuum regions was partially based on the spectral index definitions in \citet{Harbeck2003}, i.e., $3600-3630$ \AA, $3894-3910$ \AA, $4055-4080$ \AA, $4240-4280$ \AA, $4390-4460$ \AA, $4470-4500$ \AA. Since these continuum regions are generally unaffected by strong molecular features, synthetic spectra with solar-scaled abundances are sufficient for our calibration purpose. Finally, the calibrated spectra of the five observation blocks were combined. We measured the spectral indices of CN3839 and CH4300 on the combined calibrated spectrum, according to the index definition of \citet{Harbeck2003}. 
\begin{equation}
{\rm CN3839}=-2.5\log\frac{F_{3861-3884}}{F_{3894-3910}}
\end{equation}
\begin{equation}
{\rm CH4300}=-2.5\log\frac{F_{4285-4315}}{0.5F_{4240-4280}+0.5F_{4390-4460}}
\end{equation}
where $F_{X-Y}$ is the summed spectral flux from X to Y \AA. An example of the calibrated spectra can be found in Figure \ref{fig:10}.

\begin{table*}
\caption{Properties of Observed RGB Stars.}              
\label{tab:index}      
\centering                                      
\begin{tabular}{c c c c c c c c c c c c}         
\hline\hline  
\# & NSPEC$^{a}$& RA &  DEC & T$_{\rm eff}^{b}$ & T$_{\rm eff}$ Err & CN3839 & CN3839 Err & CH4300  & CH4300 Err & B11 RV  & $r$ mag\\
&  & (deg) &  (deg) & (K) & (K)  &   &   &    &   & (km/s)  &  \\
\hline 
  1&   9& 346.664032 &12.793139  &5296.6 & 123.9 & 0.115 & 0.026 &0.229 & 0.007& 24.15 &18.99\\
  2&  12& 346.677368 &12.782861  &5306.6 & 125.0 & 0.088 & 0.031 &0.201 & 0.008& 25.52 &19.08\\
  3&  14& 346.690613 &12.778083  &5421.1 & 137.5 & 0.130 & 0.032 &0.218 & 0.007& 25.91 &19.54\\
  4&  16& 346.686890 &12.775111  &5070.6 & 94.7 & 0.122 & 0.043 &0.261 & 0.017& 27.03 &18.77\\
  5&  17& 346.690247 &12.772750  &5381.8 & 133.4 & 0.097 & 0.033 &0.227 & 0.010& 23.72 &19.46\\
  6&  18& 346.702179 &12.772000  &4994.7 & 83.6 & -0.044 & 0.033 &0.236 & 0.017& 29.22 &17.32\\
  7&  19& 346.685120 &12.769500  &5206.9 & 113.0 & 0.042 & 0.028 &0.237 & 0.005& 25.58 &18.74\\
  8&  21& 346.701111 &12.762944  &5319.9 & 126.5 & 0.046 & 0.034 &0.170 & 0.010& 19.98 &18.3\\
  9&  22& 346.682587 &12.760056  &5333.4 & 128.0 & 0.082 & 0.026 &0.219 & 0.005& 25.66 &18.7\\
 10&  23& 346.674988 &12.757361  &4842.2 & 59.1 & 0.172 & 0.048 &0.350 & 0.011& 20.27 &17.68\\
 11&  26& 346.666595 &12.746444  &5333.4 & 128.0 & 0.067 & 0.021 &0.204 & 0.005& 26.44 &19.35\\
 12&  29& 346.689056 &12.739306  &5323.3 & 126.9 & 0.091 & 0.016 &0.188 & 0.004& 21.94 &19.59\\
 13&  33& 346.704742 &12.725389  &5366.1 & 131.7 & 0.111 & 0.012 &0.192 & 0.004& 28.51 &20.18\\
 14&  10& 346.708740 &12.787278  &4816.0 & 54.6 & -0.037 & 0.016 &0.280 & 0.021& 25.44 &16.66\\
\hline                                             
\end{tabular}

\raggedright{\hspace{0.2cm}$^a$ Star labels in Figure \ref{fig:slit}. $^b$ Photometric T$_{\rm eff}$.
}\\
\end{table*}

\subsection{Spectral Indices}
\begin{figure*}
\centering
\includegraphics [width=0.8\textwidth]{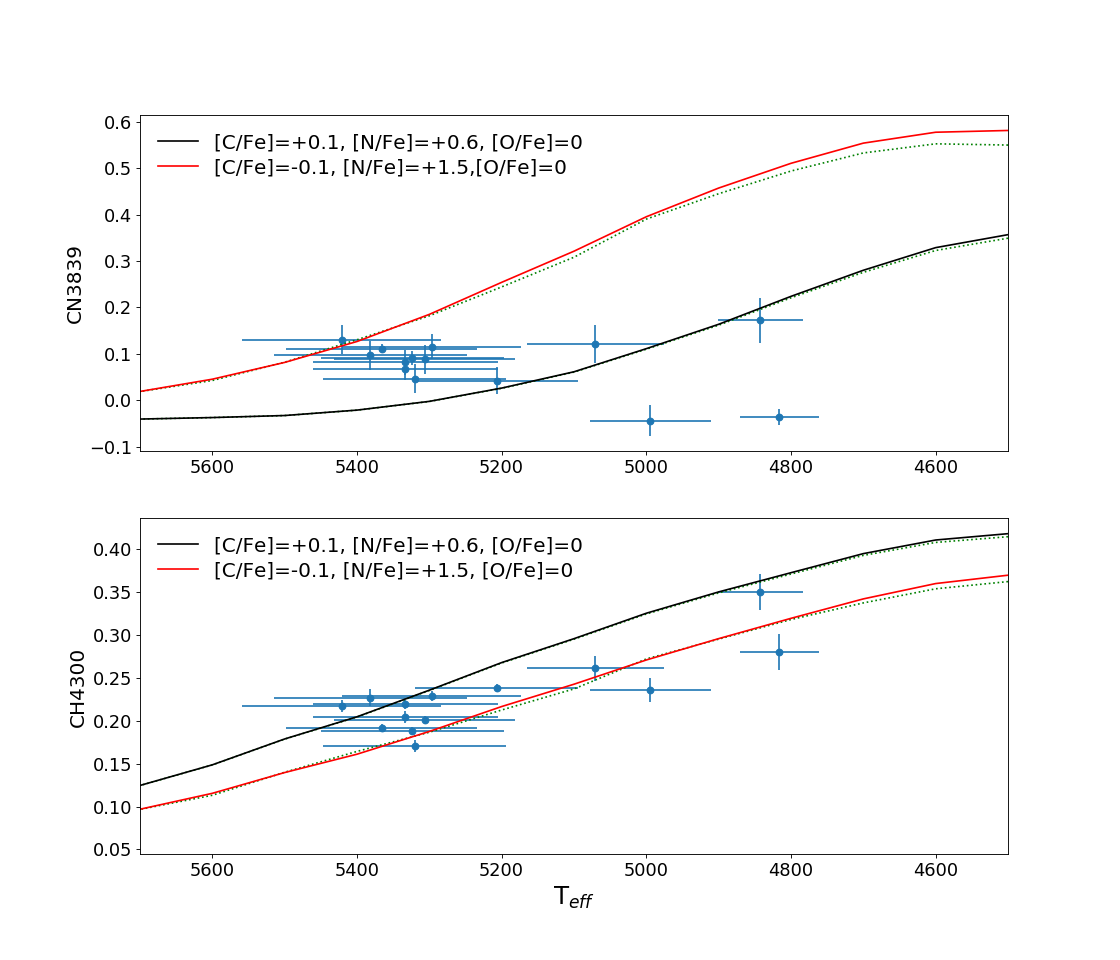} 
\caption{CN3839 and CH4142 as a function of T$_{\rm eff}$. The RGB members of Pal 13 are presented as error bars. To guide the eye, we plot stellar models with two chemical compositions: [C/Fe$]=+0.1$, [N/Fe$]=+0.6$ (black lines) and [C/Fe$]=-0.1$, [N/Fe$]=+1.5$ (red lines).  Enhanced O abundance models ([O/Fe]=$+0.4$, green dotted lines) for both chemical compositions are shown to quantify the O effect in spectral indices.  
}\label{fig:CNCH}
\end{figure*}

We estimated the uncertainties of T$_{\rm eff}$ based on the uncertainties of $(B-V)$. We adopted 0.04 mag as the uncertainty of $(B-V)$ for all the science targets, given that the $(B-V)$ color - $(g-r)$ color conversion has an uncertainty of 0.04 mag,  and the Stetson $(B-V)$ measurement uncertainties are less than 0.04 mag. The uncertainties in T$_{\rm eff}$ were propagated from $(B-V)$ using the relation between these two quantities that we inferred before. Moreover, we also estimated the uncertainties of the spectral indices by considering (1) different order polynomials when fitting the spectrum of each observation block; and (2) flux standard deviations of the calibrated spectra between five observation blocks. We plotted the spectral indices as a function of T$_{\rm eff}$ in Figure \ref{fig:CNCH}. Note that we do not include hotter SGB/MS stars in the discussion of spectral indices for several reasons: (1) The T$_{\rm eff}$ uncertainties for these hotter stars are two to three times larger than that of RGB stars; (2) It is difficult for CN and CH molecules to survive in stars with T$_{\rm eff}>5500$ K; (3) The relatively low SNR around and below 3839 \AA~may amplify the Balmer line contamination given that Balmer lines are strong for hotter stars.  The measurement spectral indices and their related uncertainties for RGB stars are listed in Table \ref{tab:index}.

For reference, we used iSpec to generate model spectra at different temperatures with two chemical compositions: (A) [C/Fe$]=+0.1$, [N/Fe$]=+0.6$ and (B) [C/Fe$]=-0.1$, [N/Fe$]=+1.5$. The model $ \log(g)$ are inferred from the above-mentioned PARSEC isochrone at different temperatures, while model [Fe/H] are set to $-1.6$.  As C, N, and O are involved in the CNO nucleosynthetic cycles, and also connected to each other in the formation of CN, CH, CO, and OH molecules, there is concern that O abundance may affect model spectra.  Therefore, we compared CN and CH spectral indices for two sub-models: [O/Fe$]=0$ and $+0.4$ (solid and dashed lines in Figure \ref{fig:CNCH}, respectively). Both CN and CH spectral indices slightly decrease as O abundance increases, since more O means more C is locked in CO, and less C to form CH and CN   \citep{Simpson2019}. However, this difference is almost negligible inside the discussed T$_{\rm eff}$ range: adopting [O/Fe$]=0$ or $+0.4$ does not affect the following discussion of CN and CH indices. 

We measured the spectral indices on the model spectra, and plotted them in Figure \ref{fig:CNCH} as a guide to the eye (solid lines). We chose these two chemical compositions because (1) they agree with the statement that C-N are generally anti-correlated in GCs \citep[e.g.,][]{Meszaros2020}; and (2) they better cover observation data points.  The CH index is mainly determined by [C/Fe], but not [N/Fe]. In the lower panel of Figure \ref{fig:CNCH}, we find that most of the observation data points are located between or very close to the two model lines of [C/Fe$]=-0.1$ and $+0.1$, which suggests that [C/Fe] of RGB stars should vary around $-0.1<$[C/Fe]$< +0.1$. On the other hand, The CN index is affected by the C and N abundances together, but luckily, we have decoded the C abundances from the CH index, so identifying the proper N abundances is feasible. In the upper panel of Figure \ref{fig:CNCH},we find that most of the observation data points lie between or very close to the model lines of (A) and (B) chemical compositions, except two cool stars with CN3839$< 0$. Since these two models are used for illustrative purpose, this model coverage is sufficient for our following discussion. Obviously, [N/Fe] show significant variation, which may be larger than $1.5-0.6=0.9$ dex. 
As stars ascend the RGB, [N/Fe] increase and [C/Fe] decrease due to first dredge-up and extra-mixing\cite[e.g.,][]{Iben1967,Gratton2000,Charbonnel2007,Charbonnel2010}, which leads to higher [N/Fe] but lower [C/Fe] at lower temperatures. However, we do not see such a trend in Figure \ref{fig:CNCH}: [C/Fe] vary round $-0.1$ and $0.1$ without obvious trend, while [N/Fe] are higher in high T$_{\rm eff}$ RGB stars. 
 As our RGB stars have past the first dredge-up (at the bottom of RGB) but before the RGB bump (Figure \ref{fig:CMD}) where extra-mixing begins to occur\footnote{Except the coolest and brightest star, but it does not show large CN spectral index, nor large [N/Fe] value in the following discussion.}, it is probable that all RGB stars are affected approximately equally from the first dredge up. 
As a result, the aforementioned processes do not significantly change the internal [N/Fe] variation among RGB stars.
Therefore, our observations indicate the presence of MPs in Pal 13, which may be the lowest mass, old GCs showing a MP signal so far discovery.

\subsection{C and N Abundances}
\begin{table}
\caption{C and N Abundances of Observed RGB Stars.}              
\label{tab:abun}      
\centering                                      
\begin{tabular}{c c c c c}         
\hline\hline  
\#   & [C/Fe] & [N/Fe] & [C/Fe] Err  & [N/Fe] Err \\
\hline 
  1 & 0.07  & 1.12  &   0.03  &  0.09  \\ 
  2 & -0.03  & 1.14  &   0.04  &  0.12  \\ 
  3 & 0.17  & 1.27  &   0.03  &  0.11  \\ 
  4 & -0.06  & 0.90  &   0.06  &  0.16  \\ 
  5 & 0.17  & 1.12  &   0.04  &  0.12  \\ 
  6 & -0.24  & 0.13  &   0.07  &  0.25  \\ 
  7 & -0.01  & 0.77  &   0.02  &  0.13  \\ 
  8 & -0.15  & 1.11  &   0.05  &  0.18  \\ 
  9 & 0.09  & 1.06  &   0.02  &  0.10  \\ 
 10 & 0.05  & 0.57  &   0.04  &  0.13  \\ 
 11 & 0.03  & 1.08  &   0.02  &  0.09  \\ 
 12 & -0.06  & 1.24  &   0.02  &  0.07  \\ 
 13 & 0.01  & 1.31  &   0.02  &  0.04  \\ 
 14 & -0.24  & 0.03  &   0.08  &  0.09  \\ 
\hline                                             
\end{tabular}
\end{table}

\begin{figure*}
\centering
\includegraphics [width=0.8\textwidth]{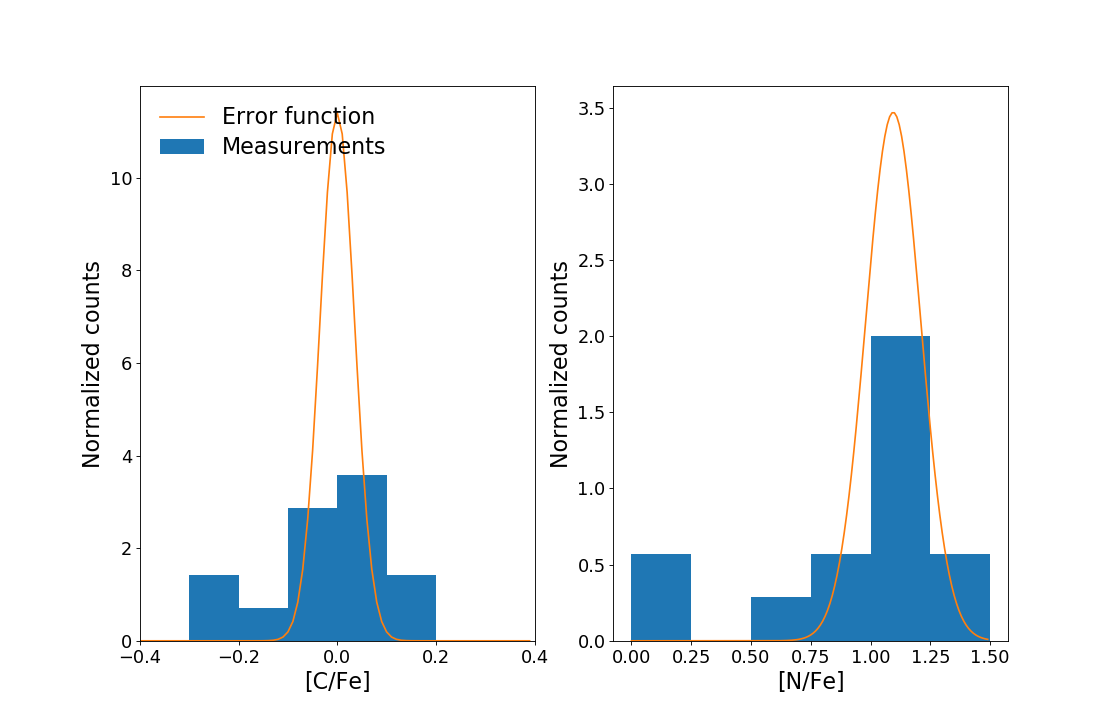} 
\caption{The distributions of [C/Fe] and [N/Fe] are shown as filled histograms in the left and right panels, respectively. The Gaussian distribution function in each panel is plotted such that the center represents the median of the measurements, and the 1-sigma represents the average measurement errors. 
}\label{fig:CN}
\end{figure*}

To further our discussion about MPs, we derived C and N abundances by comparing the observed spectral indices to stellar atmosphere models. First, we used iSpec to generate a grid of model spectra with [C/Fe] between -0.3 and 0.3 and [N/Fe] between 0 and 1.8 with a step size of 0.1 dex for each star. For each model spectrum, we measured the CN3839 and CH4300 spectral indices. Then we interpolated between model grid to find the best match to our observed spectral indices. To estimate the measurement uncertainties in C and N abundances propagated from spectral indices, we ran 1000 interpolation for each star, taking into account the uncertainties in the input data considering 1$\sigma$ variations in a Gaussian Monte Carlo approach. The means and standard deviations (Errors) of these measurements for each star are given in Table \ref{tab:abun}. Figure \ref{fig:CN} shows the distributions of [C/Fe] and [N/Fe]. To compared with the measurement uncertainties, we also show a Gaussian distribution function in each panel, whose center is the median of the measurements of all RGB stars, and sigma is their average errors. We clearly see that [C/Fe] and [N/Fe] distributions are wider than Gaussian error functions. Specifically, 5 and 3 out of 14 measurements are beyond 3$\sigma$ uncertainties for C and N abundances, respectively.  To further quantify the difference between Gaussian error function and derived measurement distribution, we employ the two-sample Kolmogorov-Smirnov (K-S) test. The two-sample K-S test is a nonparametric hypothesis test that evaluates the difference between the cumulative distribution functions (cdfs) of the two sample data vectors over the range of x in each data set. The main idea is to find the maximum difference (D-value) between two cdfs. Based on the D-value and the sample size, a p-value is calculated, which indicates if one can accept the hypothesis that two data samples are drawn from the same parent sample. We find that the probability (p-value) of drawing the Gaussian error function and the derived measurements from the same parent sample is less than 0.003 (0.0003) for [C/Fe] ([N/Fe]). Thus, we can safely reject the hypothesis that the variations in derived measurements are purely caused by errors. Our derived C and N abundances verify the existence of MPs.

\section{Discussion}
\label{sect:dis}

\subsection{The evolution of Pal 13}
\label{subsect:evo}
Recently, our knowledge of the MW formation and evolution has been revolutionized by the massive mount of data products from the Gaia mission and large spectroscopic surveys. Several dwarf galaxies are suggested to have been accreted by the MW since its formation \citep[e.g.,][]{Helmi2018,Belokurov2018, Myeong2019}. It is expected that GCs formed in these dwarf galaxies mix up with GCs formed $in-situ$, i.e. in the main MW progenitor, to form the current GC system. This formation history can be revealed by precise kinematics and dynamical modeling. \citet{Massari2019} placed Pal 13 into a group of GCs related to a dwarf galaxy, named Sequoia, which was likely accreted 9 Gyr ago \citep{Myeong2019}. The existence of MPs in Pal 13 supports the statement that this phenomenon is not unique in Galactic GCs formed $in-situ$ \citep[e.g.,][]{Li2019,Milone2020}. Besides its proposed accreted origin, Pal 13 is suggested to be experiencing tidal stripping \citep{Yepez2019, Piatti2020}. \citet{Hamren2013} found that Pal 13 
has lost a considerable amount of mass, which is related to its low present-day cluster mass. Recently, the discovery of tidal tails in Pal 13 was reported by \citet{Shipp2020}. 
Using the RR Lyrae stars, these authors estimated the initial luminosity to be L$_{V}=5.1^{+9.7}_{-3.4} \times 10^3$ L$_{\odot}$, which is significantly larger than the current luminosity estimated by B11,  L$_{V}=1.1^{+0.5}_{-0.3} \times 10^3$ L$_{\odot}$.  Due to its large distance, several detailed features of Pal 13 have been discovered just recently \citep[e.g.][]{Piatti2020, Shipp2020} and more are still waiting for further investigation, the estimation of total mass loss or initial mass of Pal 13 should be illuminative for future studies. Besides Pal 13, several remote, low mass GCs, including Whiting 1 and Eridanus GC, also show evidence of tidal tails or extra-tidal structures\citep{Carballo-Bello2014,Myeong2017}. These features agree with the predictions of dynamical evolution of low mass GCs: it is difficult to keep stars in a shallow potential well. The relatively large stellar mass lost from  low mass GCs also complicates the discussion of MPs and especially of the lower mass limit needed to maintain MPs. On the other hand, the N-rich stars located in these GCs are lost to the field, contributing to the rare N-rich field stars \citep[e.g.,][]{Martell2010, Fernandez-Trincado2017a, Tang2019, FT2019c, Tang2020}.


 \begin{figure*}
\centering
\includegraphics [width=0.95\textwidth]{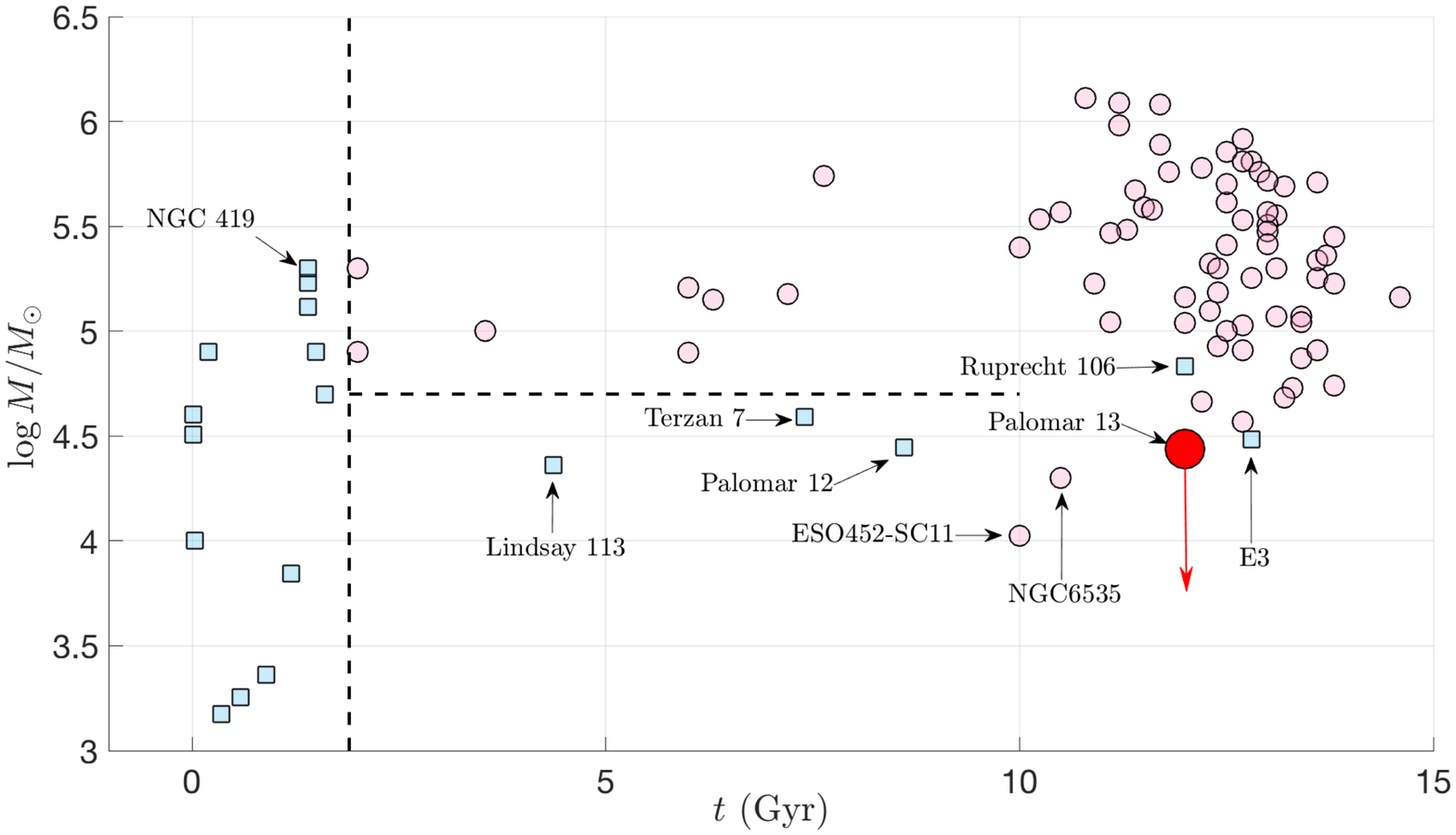} 
\caption{Age-mass parameter plane. This is an updated version of Figure 12 of \citet{Li2019}, where most of the data come from \citet{Krause2016}. We have updated the stellar masses of low mass, old GCs using \citet{Baumgardt2018}. We denote the possible overestimation of Pal 13 mass in  \citet{Baumgardt2018} with an red arrow (see text for more detailed). GCs with MP detection are labeled as pink dots, while GCs without MP detection are labeled as blue squares.  The vertical and horizontal dashed lines indicate suggested age and mass boundaries, which may define the presence of MPs.   
}\label{fig:mp}
\end{figure*}

\subsection{Implication on MP investigations}
The major goal of this work is to examine the possible MP phenomenon in low mass GCs, and provide physical constraints to various MP models. GC mass thus deserves further discussion. In Section \ref{sect:intro}, we listed the masses of several low mass GCs, including Pal 13. However, comparing GC masses derived by different methods may introduce large systematic errors. To put them on the same scale, we refer to \citet{Baumgardt2018} for GC masses derived by fitting a large set of N-body simulations to their velocity dispersion and surface density profiles. They suggested Pal 13 mass $=2.74\pm1.45\times 10^4$ M$_{\odot}$, E3 mass $=3.06\pm1.69\times 10^4$ M$_{\odot}$, ESO452-SC11 mass $=1.06\pm0.57\times 10^4$ M$_{\odot}$, and NGC 6535 mass $=2.0\pm0.56\times 10^4$ M$_{\odot}$. Though Pal 13 is not the GC with the lowest mass among them at first glance, these authors cautioned that their resulting $M/L$ ratio ($\sim10$) is significantly larger than other GCs in their study, and also much larger than the $M/L$ ratio estimated by B11 ($\sim2$). This indicates that undetected binaries might cause much larger stellar mass estimation for Pal 13 in the work of \citet{Baumgardt2018}. We note that B11 found the dynamical mass inside the half-light radius to be $1.3^{+2.7}_{-1.3} \times 10^3$ M$_{\odot}$. The large uncertainties in present-day mass estimation for low mass GCs also make estimating precise mass threshold for MPs very difficult. 

Zooming out from the four low mass GCs that we mentioned above, we further investigate the current picture of MPs in the age-mass parameter plane (Figure \ref{fig:mp}, updated version of Figure 12 of \citealt{Li2019}). For old GCs (age $>10$ Gyr), MPs are prominent. Our discovery of MPs in Pal 13 agrees with other two old low mass GCs, NGC 6535 and ESO452-SC11. From the currently available data, we suggest MPs can exist for GC mass around $1-3\times 10^4$ M$_{\odot}$, given their uncertainties in mass estimation. In that respect, the lack of MPs in E3 \citep{Salinas2015, Monaco2018} should be considered as an outlier, comparable to the similar case in Rup 106 \citep{Villanova2013, Dotter2018}. It is possible that other unknown factors, besides mass and age, affect the formation of MPs in these outliers. One possible option is mass loss, which is prominent in low mass GCs. 

Looking back at the young and intermediate GCs that have been found in the Magellanic Clouds or related to the accreted dwarf galaxies in our MW, we see that age and mass both have significant impact on the existence of MPs \citep[e.g.,][]{Martocchia2017, Martocchia2018}. 
It is interesting to note that a low mass, intermediate age, LMC (Large Magellanic Cloud) cluster, Lindsay 113 \citep{Li2019W}, harbors no MPs. Since Lindsay 113 has a comparable mass ($2.3\times 10^4$ M$_{\odot}$) to the four low mass, old GCs, Figure \ref{fig:mp} implicates age could be one of the key factors to the presence of MPs at face value. 
If age is the determining factor for the presence of MPs, new stellar models may be required. \citet{Bastian2018} suggested that magnetic braking may play a role on the formation of MPs, as they found that only star clusters with no fast rotating stars (thus their evolved stars are magnetically braked) show the presence of MPs. However, \citet{Li2020} carefully studied the population of late-type MS stars in the cluster NGC 419. They reported that even for these magetically braked MS stars, no evidence of MPs is found. 

Besides age, initial GC mass has been suggested to be related to the presence of MPs \citep[e.g.,][]{Carretta2010b, Conroy2011, Milone2020}. Specifically, initial GC mass is determined by the current GC mass, mass loss during early GC evolution ($\sim 10^8$ yr, including accretion from the ambient interstellar medium, ISM), and mass loss after GC star formation ($\sim 10^{10}$ yr).  Assuming that the second generation stars are formed in the contaminated gas enriched by the ejecta of first generation stars, normal star formation scenarios required the first generation to have been substantially more massive (by a factor of 10-100) than its present mass \citep{DErcole2008}. In this case, GCs must lose 90\% or more of its initial mass during its early evolution. To resolve the discrepancy between mass loss required by star formation scenarios and mass loss inferred from observed massive stellar remnants, \citet{Conroy2011} presented a novel model, e.g., (1) invoking significant accretion during the development of the GC gas supply, and (2) delaying star formation for several $10^8$ yr due to the high Lyman-Werner photon flux density. One of their solutions gives threshold mass consistent with what is found in LMC clusters ($\sim 10^4$ M$_{\odot}$, the horizontal dashed line of Figure \ref{fig:mp}).  

Estimating initial mass for old GCs becomes more complicated because mass loss after GC star formation ($\sim 10^{10}$ yr) may be significant compared to its initial GC mass. In this sense, comparing two GCs with similar stellar masses, but different ages may be illustrative. Given that Lindsay 113 is only 4.5 Gyr-old, the absence of MPs in Lindsay 113 may be explained by lower initial GC mass, as its total mass loss over its lifetime is presumably lower than that of old MW GCs, assuming similar mass loss rate \citep[e.g.,][]{Fall2012}, or smaller mass loss rate in LMC star clusters \citep[e.g.,][]{Milone2020}. On the contrary, most of the old GCs should have large mass losses over their long lifetime (e.g., Pal 13), and thus large initial GC masses, where a large potential well helps maintain enriched gas and generate MPs in their early days \citep[e.g.,][]{Conroy2011}. If the mass losses of old GCs are particularly small for some reasons, e.g., weak interaction with the MW, then their smaller initial GC masses may cause the absence of MPs. 

In this sense, if the MP threshold mass dramatically increase for extremely young clusters (e.g., younger than 100 Myr or so), the absence of MPs among young massive clusters may be easily fitted into this scenario. Therefore, it is essential to search for MPs in extremely young massive clusters ($<100$ Myr, $> 10^6$ M$_{\odot}$). Such super massive and extremely young clusters have not been found in our Galaxy or the Magellanic Clouds, but it is possible to find them in nearby galaxies. However, integrated light of several extremely young and super massive clusters do not show clear sign of MPs \citep{Lardo2017, Bastian2020}. This may indicate that MPs arise after a certain evolution age (e.g., $\sim 2$ Gyr).

\section{Conclusion}
\label{sect:con}

In this work, we addressed the yet-unknown MP phenomenon by searching for its presence in the low mass, old GC, Pal 13. We obtained $U$ band low-resolution spectra of 30 GC members using OSIRIS/Multi-object spectrograph mounted at the Gran Telescopio Canarias. After measuring the CN and CH related spectral indices, we detect clear nitrogen variation among 14 RGB stars. This indicates MPs exist in the GC Pal 13. This GC recently attracts attention because of its revealed peculiarity under deep photometric observations, e.g., high binary fraction \citep{Clark2004}, experiencing tidal dissolution \citep{Piatti2020}, and the discovery of tidal tails \citep{Shipp2020}. Our novel discovery of MPs in Palomar 13 is consistent with the suggested large mass loss, assuming that MPs are preferentially formed in large initial mass GCs \citep{Conroy2011}. Our study advocates the importance of mass loss when investigating MPs in low mass, old GCs. Assuming large mass loss after GC star formation ($\sim 10^{10}$ yr), MPs may commonly exist in low mass ($1-3\times 10^4$ M$_{\odot}$), old GCs. Further investigation of MPs on other low mass, old GCs will allow us to verify this statement. Using UV filters on board of Hubble Space Telescope and the up-coming Chinese Space Station Telescope, we will be able to explore MP signals in the main sequence for low mass GCs. The next-generation telescopes (e.g., Thirty Meter Telescope, Extremely Large Telescope) will provide high-resolution spectra with sufficient SNR for a large number of member stars in low mass, old GCs, which will enlighten the future discussion of this topic.


\section{acknowledgments}
We thank Divakara Mayya and Gabriel G\'omez Velarde for helpful discussions. We thank the anonymous referee for insightful comments. 
B.T. gratefully acknowledges support from National Natural Science Foundation of China under grant No. U1931102 and support from the hundred-talent project of Sun Yat-sen University. 
Y. W. was supported by the National Natural Science Foundation of China under grants 11803048, and by 
the Special Research Assistant Fundation Project of Chinese Academy of Sciences.
C. L. was supported by the National Natural Science Foundation of China through grants 12037090, Guangdong Major Project of Basic and Applied Basic Research (Grant No. 2019B030302001) and the hundred-talent project of Sun Yat-sen University.
D.G. gratefully acknowledges support from the Chilean Centro de Excelencia en Astrof\'isica
y Tecnolog\'ias Afines (CATA) BASAL grant AFB-170002.
D.G. also acknowledges financial support from the Direcci\'on de Investigaci\'on y Desarrollo de
la Universidad de La Serena through the Programa de Incentivo a la Investigaci\'on de
Acad\'emicos (PIA-DIDULS).

This work is based on observations made with the Gran Telescopio Canarias (GTC), installed in the Spanish Observatorio del Roque de los Muchachos of the Instituto de Astrofisica de Canarias, in the island of La Palma.
This work uses results from the European Space Agency (ESA) space mission Gaia. Gaia data are being processed by the Gaia Data Processing and Analysis Consortium (DPAC). Funding for the DPAC is provided by national institutions, in particular the institutions participating in the Gaia MultiLateral Agreement (MLA). 

\bibliographystyle{aasjournal}
\bibliography{gc,survey,cnchcor}

\label{lastpage}

\end{document}